\documentstyle[12pt,twoside, epsf]{article}

\def\picill#1by#2(#3)
{\vbox to #2
{\hrule width #1 height 0pt depth 0pt
\vfill\epsffile{#3}}}

\let \ttorg \tt \def \tt{\ttorg \obeyspaces}

\begin{document}

\date{}

\large{}

\title{\bf Non-Commutative Worlds - A Summary}

\author{Louis H. Kauffman \\
  Department of Mathematics, Statistics and Computer Science \\
  University of Illinois at Chicago \\
  851 South Morgan Street\\
  Chicago, IL, 60607-7045}

\maketitle
  
\thispagestyle{empty}

\section{Introduction to Non-Commutative Worlds}
Aspects of gauge
theory, Hamiltonian mechanics and quantum mechanics arise naturally in the mathematics of a non-commutative
framework for calculus and differential geometry. This paper consists in two sections. This first section sketches
our results in this domain in general. The second section gives a derivation of a generalization of the Feynman-Dyson 
derivation of electromagnetism using our non-commutative context and using diagrammatic techniques.
The first section is based on the paper \cite{NCW}. The second section is a new approach to issues in \cite{NCW}.
\bigbreak

Constructions are performed in a Lie algebra $\cal A.$   
One may take $\cal A$ to be a specific matrix Lie algebra, or abstract Lie algebra.
If $\cal A$ is taken to be an abstract Lie algebra, then it is convenient to use the universal
enveloping algebra so that the Lie product can be expressed as a commutator. In making general constructions of
operators satisfying certain relations, it is understood that one can always begin with a free algebra and make
a quotient algebra where the relations are satisfied.
\bigbreak

On $\cal A,$ a variant of calculus  is built by
defining derivations as commutators (or more generally as Lie products). For a fixed $N$ in $\cal A$ one defines
$$\nabla_N : \cal A \longrightarrow \cal A$$ by the formula
$$\nabla_{N} F = [F, N] = FN - NF.$$
$\nabla_N$ is a derivation satisfying the Leibniz rule.  
$$\nabla_{N}(FG) = \nabla_{N}(F)G + F\nabla_{N}(G).$$
\bigbreak

There are many motivations for replacing derivatives by commutators. If $f(x)$ denotes (say) a function of a real variable $x,$
and $\tilde{f}(x) = f(x+h)$ for
a fixed increment $h,$ define the {\em discrete derivative} $Df$ by the formula $Df = (\tilde{f} - f)/h,$ and find that
the Leibniz rule is not satisfied. One has the basic formula for the discrete derivative
of a product: $$D(fg) = D(f)g + \tilde{f}D(g).$$
Correct this deviation from the Leibniz rule by introducing a new non-commutative operator $J$ with the property that 
$$fJ = J\tilde{f}.$$ Define a new discrete derivative in an extended non-commutative algebra by the formula
$$\nabla(f) = JD(f).$$ It follows at once that 
$$\nabla(fg) = JD(f)g + J\tilde{f}D(g) = JD(f)g + fJD(g) = \nabla(f)g + f\nabla(g).$$
Note that $$\nabla(f) = (J\tilde{f} - Jf)/h =
(fJ-Jf)/h = [f, J/h].$$ In the extended algebra, discrete derivatives are represented by commutators, and satisfy the Leibniz rule. 
One can regard discrete calculus as a subset of non-commutative
calculus based on commutators.
\bigbreak

In $\cal A$ there are as many derivations as there are elements of the
algebra, and these derivations behave quite wildly with respect to one another. If
one takes the concept of {\em curvature} as the non-commutation of
derivations, then $\cal A$ is a highly curved world indeed. Within $\cal A$ one can build 
a tame world of derivations that mimics the
behaviour of flat coordinates in Euclidean space. The description of the
structure of $\cal A$ with respect to these flat coordinates contains many of the
equations and patterns of mathematical physics.
\bigbreak

\noindent The
flat coordinates $X_i$ satisfy the equations below with the $P_j$ chosen to represent differentiation with
respect to $X_j.$:

$$[X_{i}, X_{j}] = 0$$
$$[P_{i},P_{j}]=0$$
$$[X_{i},P_{j}] = \delta_{ij}.$$
Derivatives are represented by commutators. 
$$\partial_{i}F = \partial F/\partial X_{i} = [F, P_{i}],$$
$$\hat{\partial_{i}}F = \partial F/\partial P_{i} = [X_{i},F].$$ 
Temporal derivative is represented by commutation with a special (Hamiltonian) element $H$ of the algebra:
$$dF/dt = [F, H].$$
(For quantum mechanics, take $i\hbar dA/dt = [A, H].$)
These non-commutative coordinates are the simplest flat set of
coordinates for description of temporal phenomena in a non-commutative world.
Note:

\noindent {\bf Hamilton's Equations.} $$dP_{i}/dt = [P_{i}, H] = -[H, P_{i}] = -\partial H/\partial X_{i}$$
$$dX_{i}/dt = [X_{i}, H] = \partial H/\partial P_{i}.$$ 
These are exactly Hamilton's equations of motion. The pattern of
Hamilton's equations is built into the system.
\bigbreak

\noindent {\bf Discrete Measurement.} Consider a time series $\{X, X', X'', \cdots \}$ with commuting scalar values.
Let $$\dot{X} = \nabla X = JDX = J(X'-X)/\tau$$ where $\tau$ is an elementary time step (If $X$ denotes a times series value at time
$t$, then 
$X'$ denotes the value of the series at time $t + \tau.$). The shift operator $J$ is defined by the equation
$XJ = JX'$
where this refers to any point in the time series so that $X^{(n)}J = JX^{(n+1)}$ for any non-negative integer $n.$
Moving $J$ across a variable from left to right, corresponds to one tick of the clock. This discrete,
non-commutative time derivative satisfies the Leibniz rule. 
\bigbreak

This derivative $\nabla$ also fits a significant pattern of discrete observation. Consider the act of observing $X$ at a given time
and the act of observing (or obtaining) $DX$ at a given time. 
Since $X$ and $X'$ are ingredients in computing $(X'-X)/\tau,$ the numerical value associated with $DX,$ it is necessary  to let the
clock tick once, Thus, if one first observe
$X$ and then obtains $DX,$ the result is different (for the $X$ measurement) if one first obtains $DX,$ and then observes $X.$ In the
second case, one finds the value $X'$ instead of the value $X,$ due to the tick of the clock. 
\bigbreak

\begin{enumerate}
\item Let $\dot{X}X$ denote the sequence: observe $X$, then obtain $\dot{X}.$ 
\item Let $X\dot{X}$ denote the sequence: obtain $\dot{X}$, then observe $X.$ 
\end{enumerate}
\bigbreak

The commutator $[X, \dot{X}]$ expresses the difference between these two orders of discrete measurement.
In the simplest case, where the elements of the time series are commuting scalars, one has
$$[X,\dot{X}] = X\dot{X} - \dot{X}X =J(X'-X)^{2}/\tau.$$
Thus one can interpret the equation $$[X,\dot{X}] = Jk$$ ($k$ a constant scalar) as $$(X'-X)^{2}/\tau = k.$$ This means
that the process is a walk with spatial step $$\Delta = \pm \sqrt{k\tau}$$ where $k$ is a constant. In other words, one 
has the equation
$$k = \Delta^{2}/\tau.$$ 
This is the diffusion constant for a Brownian walk.
A walk with spatial step size  $\Delta$ and time step $\tau$ will satisfy the commutator equation above
exactly when the square of the spatial step divided by the time step remains constant. This
shows that the diffusion constant of a Brownian process is a structural property of that process, independent of considerations of
probability and continuum limits.  
\bigbreak

\noindent {\bf Heisenberg/Schr\"{o}dinger Equation.} Here is how the Heisenberg form of Schr\"{o}dinger's equation fits in
this context. Let the time shift operator be given by the equation
$J=(1 + H\Delta t/i \hbar).$ Then the non-commutative version of the discrete time derivative is expressed by the commutator
$$\nabla\psi = [\psi, J/\Delta t],$$ and we calculate
$$\nabla \psi = \psi[(1 + H \Delta t/i \hbar)/\Delta t] - 
[(1 +  H\Delta t/i \hbar)/\Delta t] \psi = [\psi, H]/i \hbar,$$
$$i \hbar \nabla \psi = [\psi, H].$$
This is exactly the Heisenberg version of the Schr\"{o}dinger equation.
\bigbreak

\noindent {\bf Dynamics and Gauge Theory.} One can take the general dynamical
equation in the form 
$$dX_{i}/dt = {\cal G}_{i}$$ where $\{ {\cal G}_{1},\cdots, {\cal G}_{d} \}$
is a collection of elements of $\cal A.$ Write ${\cal G}_{i}$
relative to the flat coordinates via ${\cal G}_{i} = P_{i} -  A_{i}.$
This is a definition of $A_{i}$ and $\partial F/\partial X_{i} = [F,P_{i}].$ The formalism of gauge theory appears
naturally. In particular, if $$\nabla_{i}(F) = [F, {\cal G}_{i}],$$ then one has
the curvature $$[\nabla_{i}, \nabla_{j}]F = [R_{ij}, F]$$
and 
$$R_{ij} = \partial_{i} A_{j} - \partial_{j} A_{i} + [A_{i}, A_{j}].$$  This is the well-known formula for the curvature of a gauge
connection. Aspects of geometry arise naturally in this context, including the Levi-Civita
connection (which is seen as a consequence of the Jacobi identity in an appropriate non-commutative world).  
\bigbreak

One can consider the consequences of the commutator $[X_{i}, \dot{X_{j}}] = g_{ij}$,
deriving that  
$$\ddot{X_{r}} = G_{r} + F_{rs}\dot{X^{s}} + \Gamma_{rst}\dot{X^{s}}\dot{X^{t}},$$
where $G_{r}$ is the analogue of a scalar field, $F_{rs}$ is the analogue of a gauge field and $\Gamma_{rst}$ is the Levi-Civita
connection associated with $g_{ij}.$
This decompositon of the acceleration is uniquely determined by the given framework.
\bigbreak

One can use this context to revisit the Feynman-Dyson derivation of electromagnetism from commutator equations, 
showing that most of the derivation is independent of any choice of commutators, but highly dependent upon the choice of definitions
of the derivatives involved. Without any assumptions about initial commutator equations, but taking the right (in some sense simplest)
definitions of the derivatives one obtains a significant generalization of the result of Feynman-Dyson.  
\bigbreak

\noindent {\bf Electromagnetic Theorem.} (See Section 2.) With the appropriate [see below] definitions of the operators, and taking
$$\nabla^{2} = \partial_{1}^{2} + \partial_{2}^{2} + \partial_{3}^{2}, \,\,\, B = \dot{X} \times \dot{X} \,\,\, \mbox{and} \,\,\, E =
\partial_{t}\dot{X}, \,\,\, \mbox{one has}$$

\begin{enumerate}
\item $\ddot{X} = E + \dot{X} \times B$
\item $\nabla \bullet B = 0$
\item $\partial_{t}B + \nabla \times E = B \times B$
\item $\partial_{t}E - \nabla \times B = (\partial_{t}^{2} - \nabla^{2})\dot{X}$
\end{enumerate}
\bigbreak

The key to the proof of this Theorem is the definition of the time derivative. This definition is as follows
$$\partial_{t}F = \dot{F} - \Sigma_{i}\dot{X_{i}}\partial_{i}(F) =  \dot{F} - \Sigma_{i} \dot{X_{i}}[F, \dot{X_{i}}]$$
for all elements or vectors of elements $F.$  The definition creates a
distinction between space and time in the non-commutative world.  A calculation ( done diagrammatically in
Figure 3) reveals that
$$\ddot{X} = \partial_{t}\dot{X} + \dot{X} \times (\dot{X} \times \dot{X}).$$
This suggests taking $E = \partial_{t}\dot{X}$ as the electric field, and $B = \dot{X} \times \dot{X}$
as the magnetic field so that the Lorentz force law 
$$\ddot{X} = E + \dot{X} \times B$$
is satisfied.
\bigbreak

\noindent 
This result is applied to produce many discrete models of the Theorem. These models show that, just as the commutator $[X, \dot{X}] =
Jk$ describes Brownian motion in one dimension, a generalization of electromagnetism describes the interaction of triples of time
series in three dimensions.
\bigbreak

\noindent {\bf Remark.} While there is a large
literature on non-commutative geometry, emanating from the idea of replacing a space by its ring of  functions, work discussed herein
is not written in that tradition. Non-commutative geometry does occur here, in the sense of geometry occuring in the context of
non-commutative algebra. Derivations are represented by commutators. There are relationships between the present work and the
traditional non-commutative geometry, but that is a subject for further exploration. In no way is this paper intended to be an
introduction to that subject. The present summary is based on
\cite{Kauff:KP,KN:QEM,KN:Dirac,KN:DG,Twist,NonCom,ST,Aspects,Boundaries,NCW} and the references cited therein.
\bigbreak

The following references in relation to non-commutative calculus are useful in 
comparing with the present approach \cite{Connes, Dimakis, Forgy, MH}. Much of the present work is the fruit of a long
series of discussions with Pierre Noyes, influenced at critical points by Tom Etter and Keith Bowden. 
Paper \cite{Mont} also works with minimal coupling for the Feynman-Dyson derivation. The first remark about
the minimal coupling occurs in the original paper by Dyson \cite{Dyson}, in the context of Poisson brackets.
The paper \cite{Hughes} is worth reading as a companion to Dyson.   It is the purpose of this summary to indicate how
non-commutative calculus can be used in foundations.
\bigbreak 
 
\section{Generalized Feynman Dyson Derivation}
In this section we assume that specific time-varying coordinate elements $X_{1},X_{2},X_{3}$ of the algebra $\cal{A}$ are given.
{\it We do not assume any commutation relations about $X_{1},X_{2},X_{3}.$}
\bigbreak

In this section we no longer avail ourselves of the commutation relations that are in back of the original
Feynman-Dyson derivation. We do take the definitions of
the derivations from that previous context. Surprisingly, the result is very similar to the one of Feynman and Dyson, as
we shall see.
\bigbreak

Here $A \times B$ is the non-commutative vector cross product:
$$(A \times B)_{k} = \Sigma_{i,j = 1}^{3} \epsilon_{ijk}A_{i}B_{j}.$$ (We will drop this summation sign
for vector cross products from now on.) 
Then, with $B = \dot{X} \times \dot{X},$ we have $$B_{k} = \epsilon_{ijk}\dot{X_{i}}\dot{X_{j}}  =
(1/2)\epsilon_{ijk}[\dot{X_{i}},\dot{X_{j}}].$$ The epsilon tensor $\epsilon_{ijk}$ is defined for the indices $\{ i,j,k \}$ ranging
from $1$ to $3,$ and is equal to 
$0$ if there is a repeated index and is ortherwise equal to the sign of the permutation of $123$ given by $ijk.$
We represent dot products and cross products in diagrammatic tensor notation as indicated in Figure 1 and Figure 2. 
In Figure 1 we indicate the epsilon tensor by a trivalent vertex. The indices of the tensor correspond to labels for the 
three edges that impinge on the vertex. The diagram is drawn in the plane, and is well-defined since the epsilon tensor is
invariant under cyclic permutation of its indices.
\bigbreak

We will define the fields $E$  and $B$ by the equations 
$$B = \dot{X} \times \dot{X} \,\,\, \mbox{and} \,\,\, E =
\partial_{t}\dot{X}.$$
We will see that $E$ and $B$ obey a 
generalization of the Maxwell Equations, and that this generalization describes specific discrete models.
The reader should note that this means that a significant part of the {\it form} of electromagnetism is
the consequence of choosing three coordinates of space, and the definitions of spatial and temporal derivatives with respect to them.
The background process that is being described is otherwise aribitrary, and yet appears to obey physical laws once these
choices are made.
\bigbreak

In this section we will use diagrammatic matrix methods to carry out the mathematics.
In general, in a diagram for matrix or tensor composition, we sum over all indices labeling any edge in the diagram that has no free
ends. Thus matrix multiplication corresponds to the connecting of edges between diagrams, and to the summation over
common indices. With this interpretation of compositions, view the first identity in Figure 1. This is a fundmental identity about
the epsilon, and corresponds to the following lemma.  
\bigbreak

 \begin{center}
$$ \picill4inby4.2in(EpsilonIdentity)  $$
{ \bf Figure 1 - Epsilon Identity} 
\end{center}  
\bigbreak

\noindent {\bf Lemma.} (View Figure 1) Let $\epsilon_{ijk}$ be the epsilon tensor taking values $0$, $1$ and $-1$ as follows: When
$ijk$ is a permuation of $123$, then $\epsilon_{ijk}$ is equal to the sign of the permutation. When $ijk$ contains a repetition from 
$\{1,2,3 \},$ then the value of epsilon is zero. 
Then $\epsilon$ satisfies the following identity in terms of the Kronecker delta. 

\begin{center}
$$ \picill4inby1in(LabeledEpsilonIdentity)  $$
\end{center}  
\bigbreak

$$\Sigma_{i} \,\epsilon_{abi}\epsilon_{cdi} =  -\delta_{ad}\delta_{bc} + \delta_{ac}\delta_{bd}.$$
\bigbreak

\noindent The proof of this identity is left to the reader. The identity itself will be referred to as the {\em epsilon identity}.
The epsilon identity is a key structure in the work of this section, and indeed in all formulas involving the vector cross product.
\bigbreak

The reader should compare the formula in this Lemma with the diagrams in Figure 1. The first two diagram are two versions of the 
Lemma. In the third diagram the labels are capitalized and refer to vectors $A,B$ and $C.$ We then see that the epsilon identity
becomes the formula $$A \times (B \times C) = (A \bullet C)B - (A \bullet B)C$$ for vectors in three-dimensional space
(with commuting coordinates, and a generalization of this identity to our non-commutative context. Refer to Figure 2 for the
diagrammatic definitions of dot and cross product of vectors. We take these definitions (with implicit order of multiplication)
in the non-commutative context.
\bigbreak

\begin{center}
$$ \picill4inby5.2in(DefiningDiff)  $$
{ \bf Figure 2 - Defining Derivatives} 
\end{center}  
\bigbreak

\noindent {\bf Remarks on the Derivatives.}
\begin{enumerate}
\item Since we do not assume that $[X_{i}, \dot{X_{j}}] = \delta_{ij},$ nor do we assume $[X_{i},X_{j}]=0,$ it will not follow that
$E$ and $B$ commute with the $X_{i}.$ 

\item We define $$\partial_{i}(F) = [F, \dot{X_{i}}],$$ and the reader should note
that, these spatial derivations are no longer flat in the sense of section 1 (nor were they in the original Feynman-Dyson derivation).
See Figure 2 for the diagrammatic version of this definition.

\item We define $\partial_{t} = \partial/\partial t$ by the equation
$$\partial_{t}F = \dot{F} - \Sigma_{i}\dot{X_{i}}\partial_{i}(F) =  \dot{F} - \Sigma_{i} \dot{X_{i}}[F, \dot{X_{i}}]$$
for all elements or vectors of elements $F.$ We take this equation as the global definition
of the temporal partial derivative, even for elements that are not commuting with the $X_{i}.$ This notion of temporal partial
derivative
$\partial_{t}$ is a least relation that we can write to describe the temporal relationship of an arbitrary non-commutative vector
$F$ and the non-commutative coordinate vector $X.$ See Figure 2 for the diagrammatic version of this definition.

\item In defining $$\partial_{t}F = \dot{F} - \Sigma_{i}\dot{X_{i}}\partial_{i}(F),$$ we
are using the definition itself to obtain a notion of the variation of $F$ with respect to time. The definition itself creates a
distinction between space and time in the non-commutative world.  

\item The reader will have no difficulty verifying the following formula:
$$\partial_{t}(FG) = \partial_{t}(F)G + F\partial_{t}(G) + \Sigma_{i}\partial_{i}(F)\partial_{i}(G).$$
This formula shows that $\partial_{t}$ does not satisfy the Leibniz rule in our non-commutative context.
This is true for the original Feynman-Dyson context, and for our generalization of it. All derivations in this theory that are defined
directly as commutators do satisfy the Leibniz rule. Thus $\partial_{t}$ is an operator in our theory that does not have a
representation as a commutator.

\item We define divergence and curl by the equations
$$\nabla \bullet B = \Sigma_{i=1}^{3} \partial_{i}(B_{i})$$ and 
$$(\nabla \times E)_{k} = \epsilon_{ijk}\partial_{i}(E_{j}).$$
See Figure 2 and Figure 4 for the diagrammatic versions of curl and divergence.
\end{enumerate}
\bigbreak

Now view Figure 3. We see from this Figure that it follows directly from the definition of 
the time derivatives (as discussed above) that 
$$\ddot{X} = \partial_{t}\dot{X} + \dot{X} \times (\dot{X} \times \dot{X}).$$

This is our motivation for defining
$$E = \partial_{t}\dot{X}$$ and 
$$B = \dot{X} \times \dot{X}.$$
With these definition in place we have
$$\ddot{X} = E + \dot{X} \times B,$$ giving an analog of the Lorentz force law for
this theory. 
\bigbreak

Just for the record, look at the following algebraic calculation for this derivative:
$$ \dot{F} = \partial_{t}F  + \Sigma_{i} \dot{X_{i}}[F, \dot{X_{i}}]$$ 
$$ = \partial_{t}F + \Sigma_{i} (\dot{X_{i}}F \dot{X_{i}} - \dot{X_{i}} \dot{X_{i}} F)$$
$$ = \partial_{t}F + \Sigma_{i} (\dot{X_{i}}F \dot{X_{i}} - \dot{X_{i}} F_{i} \dot{X}) + \dot{X_{i}} F_{i} \dot{X} - \dot{X_{i}}
\dot{X_{i}} F$$
Hence $$ \dot{F} = \partial_{t}F + \dot{X} \times F  + (\dot{X} \bullet F) \dot{X} - (\dot{X} \bullet \dot{X}) F$$
(using the epsilon identity).
Thus we have
$$\ddot{X} = \partial_{t} \dot{X} + \dot{X} \times (\dot{X} \times \dot{X})  + (\dot{X} \bullet \dot{X}) \dot{X} - (\dot{X} \bullet
\dot{X})\dot{X},$$
whence
$$\ddot{X} = \partial_{t}\dot{X} + \dot{X} \times (\dot{X} \times \dot{X}).$$
\bigbreak

In Figure 4, we give the derivation that $B$ has zero divergence.

\begin{center}
$$ \picill4inby6in(Xdoubledot)  $$
{ \bf Figure 3 - The Formula for Acceleration} 
\end{center}  
\bigbreak

\begin{center}
$$ \picill4inby5in(DivB)  $$
{ \bf Figure 4 - Divergence of $B$ } 
\end{center}  
\bigbreak

Figures 5 and 6 compute derivatives of $B$ and the Curl of $E,$ culminating in the formula
$$\partial_{t}B + \nabla \times E = B \times B.$$
In classical electromagnetism, there is no term $B \times B.$ This term is an artifact of our non-commutative context.
In discrete models, as we shall see at the end of this section, there is no escaping the effects of this term.
\bigbreak

\begin{center}
$$ \picill4inby5in(Bdot)  $$
{ \bf Figure 5 - Computing $\dot{B}$} 
\end{center}  
\bigbreak

\begin{center}
$$ \picill4inby6in(CurlE)  $$
{ \bf Figure 6 - Curl of $E$} 
\end{center}  
\bigbreak

\begin{center}
$$ \picill4inby6in(CurlB)  $$
{ \bf Figure 7 - Curl of $B$} 
\end{center}  
\bigbreak

Finally, Figure 7 gives the diagrammatic proof that 
$$\partial_{t}E - \nabla \times B = (\partial_{t}^{2} - \nabla^{2})\dot{X}.$$
This completes the proof of the Theorem below.
\bigbreak

\noindent {\bf Electromagnetic Theorem} With the above definitions of the operators, and taking
$$\nabla^{2} = \partial_{1}^{2} + \partial_{2}^{2} + \partial_{3}^{2}, \,\,\, B = \dot{X} \times \dot{X} \,\,\, \mbox{and} \,\,\, E =
\partial_{t}\dot{X} \,\,\, \mbox{we have}$$

\begin{enumerate}
\item $\ddot{X} = E + \dot{X} \times B$
\item $\nabla \bullet B = 0$
\item $\partial_{t}B + \nabla \times E = B \times B$
\item $\partial_{t}E - \nabla \times B = (\partial_{t}^{2} - \nabla^{2})\dot{X}$
\end{enumerate}
\bigbreak

\noindent {\bf Remark.} Note that this Theorem is a non-trivial generalization of the Feynman-Dyson derivation of electromagnetic 
equations. In the Feynman-Dyson case, one assumes that the commutation relations
$$[X_{i}, X_{j}] = 0$$ and 
$$[X_{i}, \dot{X_{j}}] = \delta_{ij}$$ are given, {\em and} that the principle of commutativity is assumed, so that 
if $A$ and $B$ commute with the $X_{i}$ then $A$ and $B$ commute with each other. One then can interpret $\partial_{i}$ as a 
standard derivative with $\partial_{i}(X_{j}) = \delta_{ij}.$ Furthermore, one can verify that $E_{j}$ and $B_{j}$ both commute with
the $X_{i}.$ From this it follows that $\partial_{t}(E)$ and $\partial_{t}(B)$ have standard intepretations and that $B \times B = 0.$
The above formulation of the Theorem adds the description of $E$ as $\partial_{t}(\dot{X}),$ a non-standard use of 
$\partial_{t}$ in the original context of Feyman-Dyson, where $\partial_{t}$ would only be defined for those $A$ that commute with 
$X_{i}.$ In the same vein, the last formula $\partial_{t}E - \nabla \times B = (\partial_{t}^{2} - \nabla^{2})\dot{X}$ gives a way
to express the remaining Maxwell Equation in the Feynman-Dyson context.
\bigbreak

\noindent {\bf Remark.} Note the role played by the epsilon tensor $\epsilon_{ijk}$ throughout the construction of 
generalized electromagnetism in this section. The epsilon tensor is the structure constant for the Lie algebra of the rotation
group $SO(3).$ If we replace the epsilon tensor by a structure constant
$f_{ijk}$ for a Lie algebra ${\cal G}$of dimension $d$ such that the tensor is invariant under cyclic permutation ($f_{ijk} =
f_{kij}$), then most of the  work in this section will go over to that context. We would then have $d$ operator/variables $X_1,
\cdots X_d$ and a generalized  cross product defined on vectors of length $d$ by the equation
$$(A \times B)_{k} = f_{ijk}A_{i}B_{j}.$$
The Jacobi identity for the Lie algebra ${\cal G}$ implies that this cross product will satisfy
$$A \times (B \times C) = (A \times B) \times C + [B \times (A ] \times C)$$
where $$([B \times (A ] \times C)_{r} = f_{klr}f_{ijk}A_{i}B_{k}C_{j}.$$ This extension of the Jacobi identity 
holds as well for the case of non-commutative cross product defined by the epsilon tensor.
It is therefore of interest to explore the structure of generalized non-commutative electromagnetism over other Lie algebras
(in the above sense). This will be the subject of another paper.
\bigbreak

\subsection{Discrete Thoughts}
In the hypotheses of the Electromagnetic Theorem, we are free to take any non-commutative world, and
the Electromagnetic Theorem will
satisfied in that world. For example, we can take each $X_{i}$ to be an arbitary time series of real or complex numbers, or
bitstrings of zeroes and ones. The global time derivative is defined by $$\dot{F} = J(F' - F) = [F, J],$$ where $FJ =
JF'.$ This is the non-commutative discrete context discussed in sections 1. We will write
$$\dot{F} = J\Delta(F)$$ where $\Delta(F)$ denotes the classical discrete derivative
$$\Delta(F) = F' -F.$$ 
With this interpretation
$X$ is a vector with three real or complex coordinates at each time, and  
$$B = \dot{X} \times \dot{X} = J^{2}\Delta(X') \times \Delta(X)$$ while
$$E = \ddot{X} - \dot{X} \times (\dot{X} \times \dot{X}) = J^{2}\Delta^{2}(X) - J^{3} \Delta(X'') \times ( \Delta(X') \times
\Delta(X)).$$ Note how the non-commutative vector cross products are composed through time shifts in this context of temporal
sequences of scalars. The advantage of the generalization now becomes apparent. We can create very simple models of generalized
electromagnetism with only the simplest of discrete materials. In the case of the model in terms of triples of time series, the 
generalized electromagnetic theory is a theory of measurements of the time series whose key quantities are
$$\Delta(X') \times \Delta(X)$$ and 
$$\Delta(X'') \times (\Delta(X') \times \Delta(X)).$$
\bigbreak

It is worth noting the forms of the basic derivations in this model. We have, assuming that $F$ is a commuting scalar (or vector of
scalars) and taking $\Delta_{i} = X_{i}' - X_{i},$
$$\partial_{i}(F) = [F, \dot{X_{i}}] =[F, J\Delta_{i}] = FJ\Delta_{i} - J\Delta_{i}F = J(F'\Delta_{i} - \Delta_{i}F)
= \dot{F}\Delta_{i}$$  and for the temporal derivative we have
$$\partial_{t}F = J[1 - J \Delta' \bullet \Delta]\Delta(F)$$ where
$\Delta = (\Delta_{1}, \Delta_{2}, \Delta_{3}).$
\bigbreak

\noindent {\bf Acknowledgement.} Most of this effort was sponsored by the Defense
Advanced Research Projects Agency (DARPA) and Air Force Research Laboratory, Air
Force Materiel Command, USAF, under agreement F30602-01-2-05022.  The U.S. Government is authorized to reproduce and distribute
reprints for Government purposes notwithstanding any copyright annotations thereon. The
views and conclusions contained herein are those of the authors and should not be
interpreted as necessarily representing the official policies or endorsements,
either expressed or implied, of the Defense Advanced Research Projects Agency,
the Air Force Research Laboratory, or the U.S. Government. (Copyright 2005.) 
It gives the author great pleasure to acknowledge support from NSF Grant DMS-0245588 and to thank
Pierre Noyes and Keith Bowden for continuing conversations related to the contents of this paper.
\bigbreak


\begin{thebibliography}{99}

\bibitem{Dyson}
Dyson, F. J. [1990], Feynman's proof of the Maxwell Equations, {\em Am. J. Phys.} 58 (3), March 1990,
209-211.


\bibitem{Connes} Connes,Alain [1990],  {\em Non-commutative Geometry}
Academic Press.


\bibitem{Dimakis}
Dimakis, A. and M\"{u}ller-Hoissen [1992], F., Quantum mechanics on a lattice and q-deformations, 
{\em Phys. Lett.} 295B, p.242.


\bibitem{Forgy}
Forgy,Eric A. [2002]  Differential geometry in computational electromagnetics, 
PhD Thesis, UIUC.

 
\bibitem{Hughes}
Hughes, R. J. [1992], On Feynman's proof of the Maxwell Equations, {\em Am. J. Phys.} 60, (4),
April 1992, 301-306.
 

\bibitem{Kauff:KP} Kauffman,Louis H.[1991,1994], {\em Knots and Physics,}
World Scientific Pub.


\bibitem{KN:QEM} Kauffman,Louis H. and Noyes,H. Pierre [1996], Discrete
Physics and the Derivation of Electromagnetism from the formalism of
Quantum Mechanics, {\em Proc. of the Royal Soc. Lond. A}, {\bf 452}, pp.
81-95. 

\bibitem{KN:Dirac} Kauffman,Louis H. and Noyes,H. Pierre [1996], Discrete
Physics and the Dirac Equation, {\em Physics Letters A}, 218 ,pp.
139-146. 

\bibitem{KN:DG} Kauffman,Louis H. and Noyes,H.Pierre (In preparation)

\bibitem{Twist} Kauffman, Louis H.[1996], Quantum electrodynamic
birdtracks, {\em Twistor Newsletter Number 41} 

\bibitem{NonCom} Kauffman, Louis H. [1998], Noncommutativity and discrete
physics, {\em Physica D } 120 (1998), 125-138.

\bibitem{ST} Kauffman, Louis H. [1998], Space and time in discrete physics, 
{\em Intl. J. Gen. Syst.} Vol. 27, Nos. 1-3, 241-273.

\bibitem{Aspects}
Kauffman, Louis H. [1999], A non-commutative approach to discrete physics, 
in {\em Aspects II - Proceedings of ANPA 20}, 215-238.

\bibitem{Boundaries}
Kauffman, Louis H. [2003], Non-commutative calculus and discrete physics,
in {\em Boundaries- Scientific Aspects of ANPA 24}, 73-128.

\bibitem{NCW}
Kauffman, Louis H. [2004], Non-commutative worlds, {\em New Journal of Physics 6}, 2-46.

\bibitem{Mont}
Montesinos, M. and Perez-Lorenzana, A., [1999], Minimal coupling and Feynman's proof,
arXiv:quant-phy/9810088 v2 17 Sep 1999.

\bibitem{MH}
M\"{u}ller-Hoissen,Folkert [1998],  Introduction to non-commutative geometry of commutative algebras and 
applications in physics, in {\em Proceedings of the 2nd Mexican School on Gravitation and Mathematical
Physics}, Kostanz (1998) http://kaluza.physik.uni-konstanz.de/2MS/mh/mh.html.


\bibitem{Tanimura} Tanimura,Shogo [1992], Relativistic generalization and
extension to the non-Abelian gauge theory of Feynman's proof of the
Maxwell equations, {\em Annals of Physics, vol. 220}, pp. 229-247. 


\end{thebibliography}
\end{document}